\documentclass[notitlepage]{article}[11pt]

\makeatletter
\def\@maketitle{%
  \newpage
  \begin{center}%
  \let \footnote \thanks
    {\LARGE \@title \par}%
    {\large
      \lineskip .5em%
      \begin{tabular}[t]{c}%
        \@author
      \end{tabular}\par}%
    \vskip 1em%
    {\large \@date}%
  \end{center}%
  \par }
\makeatother

\usepackage{pdfpages}
\usepackage{pgffor}
\usepackage{hyperref}
\usepackage{xr-hyper}
\usepackage{xr}
\usepackage[top=1in, bottom=1in, left=1in, right=1in]{geometry}
\usepackage{comment}	
\usepackage{amsmath}
\usepackage{amsthm}
\usepackage{amssymb}
\usepackage{graphicx}
\usepackage{mathtools}
\usepackage{cases}
\usepackage{bbm}
\usepackage{authblk}
\usepackage{caption}
\usepackage{float}
\usepackage[nodayofweek,level]{datetime}
\usepackage[british]{babel}
\usepackage[round]{natbib}
\usepackage[caption = false]{subfig}
\allowdisplaybreaks

\title{\Large{TiMEx: A Waiting Time Model for Mutually Exclusive Cancer Alterations}}
\author[1,2]{\normalsize{Simona Constantinescu}}
\author[1,2]{\normalsize{Ewa Szczurek}}
\author[1,2]{\normalsize{Pejman Mohammadi}}
\author[3]{\normalsize{{J\"{o}rg Rahnenf\"{u}hrer}}}
\author[1,2]{\normalsize{Niko Beerenwinkel\footnote{niko.beerenwinkel@bsse.ethz.ch}}}
\affil[1]{\footnotesize{Department of Biosystems Science and Engineering, ETH Z\"{u}rich, Basel}} 
\affil[2]{\footnotesize{Swiss Institute of Bioinformatics, Basel}}
\affil[3]{\footnotesize{Faculty of Statistics, Technische Universit\"{a}t Dortmund, Dortmund}}
\date{}

\newcommand{\lamobs}{{\lambda_\text{obs}}}

\newtheorem{proposition}{Proposition}

\begin{document}

\maketitle

\section*{Abstract}
\noindent Despite recent technological advances in genomic sciences, our understanding of cancer progression and its driving genetic alterations remains incomplete. Here, we introduce TiMEx, a generative probabilistic model for detecting patterns of various degrees of mutual exclusivity across genetic alterations, which can indicate pathways involved in cancer progression. TiMEx explicitly accounts for the temporal interplay between the waiting times to alterations and the observation time. In simulation studies, we show that our model outperforms previous methods for detecting mutual exclusivity. On large-scale biological datasets, TiMEx identifies gene groups with strong functional biological relevance, while also proposing many new candidates for biological validation. TiMEx possesses several advantages over previous methods, including a novel generative probabilistic model of tumorigenesis, direct estimation of the probability of mutual exclusivity interaction, computational efficiency, as well as high sensitivity in detecting gene groups involving low-frequency alterations. The R code implemented TiMEx is available at $\href{www.cbg.bsse.ethz.ch/software/TiMEx}{\text{www.cbg.bsse.ethz.ch/software/TiMEx}}$.

\section*{Introduction}
\begin{figure}[t!]
	\centerline  {\includegraphics{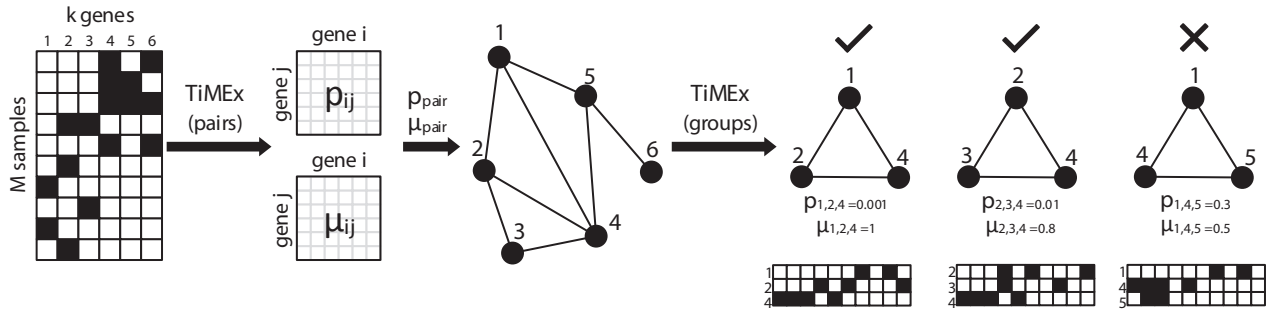}}
	\caption{\footnotesize{Overview of the TiMEx multistep procedure for detecting mutually exclusive groups of alterations in a large dataset. First, from a binary alteration matrix consisting of $M$ samples and $k$ genes, the degree of mutual exclusivity $\mu_{ij}$ and the p-value for testing $\mu_{ij} \neq 0$ against $\mu_{ij}=0$ are estimated for all gene pairs $(i,j)$. Second, candidate groups are identified as maximal cliques of genes sharing a significant minimum degree of mutual exclusivity (satisfying the thresholds $p_\text{pair}$ and $\mu_\text{pair}$ for each edge). Finally, the candidate groups are statistically tested for mutual exclusivity and the degree of mutual exclusivity corresponding to each group is estimated and tested for significance. }}
     \label{fig:overview}
\end{figure}

\noindent Despite recent technological advances in genomic sciences, our understanding of cancer progression still faces fundamental challenges. To this end, new ways of interpreting the increasing amount of generated data are devised, aiming at finding biologically relevant patterns. An important example is the separation of genes into drivers, which have a selective advantage and significantly contribute to tumor progression, and passengers, which are selectively neutral and can hitchhike along with fitter clones. Even if intuitive and routinely used, identifying drivers as recurrently altered genes \citep{driversfreq} only explains tumorigenesis in a fraction of patients. Alternatively, the functional role of drivers can be assessed in the context of groups of genes, all possessing the same important function, commonly known as pathways. Once one of the group members is altered, the tumor gains a significant selective advantage. The alteration of additional group members does not further increase the selective advantage of the tumor, making genotypes with a single alteration likely the most frequent. In this case, the group of genes displays a mutually exclusive alteration pattern.
\\
\indent Current approaches for detecting mutual exclusivity are either \textit{de novo} \citep{yeang2008,ding2008,vandinTest,multidendrix,RME,ewa} or based on biological interaction networks \citep{memo}. While highly informative, the current biological knowledge is incomplete, such that limiting the search space to known biological interactions significantly reduces the detection power. Straightforward pairwise statistical tests assessing whether the number of observed double mutants is lower than expected by chance have also been employed, followed by identifying groups as maximal cliques \citep{yeang2008,memo}. The Dendrix tool \citep{vandinTest} performs a Markov Chain Monte Carlo sampling for group structure search and then a permutation test for finding sets of genes with both high coverage and high exclusivity. Its limitation of finding the single main pathway per dataset was addressed by Multidendrix \citep{multidendrix}, a follow-up tool which simultaneously identifies multiple driver pathways via an integer linear programming approach. Finally, \cite{ewa} propose muex, a statistical model for mutual exclusivity, where, however, the group members are required to have similar alteration frequencies. All existing approaches ignore the fact that the mutually exclusive patterns occur over time, during disease progression.
\\
\indent Here, we introduce TiMEx, a generative probabilistic model for the \textit{de novo} detection of mutual exclusivity patterns of various degrees across carcinogenic alterations. We regard tumorigenesis as a dynamic process, and base our model on the temporal interplay between the waiting times to alterations, characteristic for every gene and alteration type, and the observation time. Under the assumption of rarity of events over short time intervals, TiMEx models the alteration process for each gene as a Poisson process. The waiting times to alterations are therefore modeled as exponentially distributed variables with specific rates, which correspond to the rates of evolution for each alteration. In our modeling framework, the temporal dynamics of each alteration process progresses from the onset of cancer, corresponding to the first genetic alteration responsible for the growth of a malignant tumor, up to the observation time, corresponding to the time of the tumor biopsy. The observation time is regarded as a system failure time, and is exponentially distributed with an unknown rate. 
\\
\indent A perfectly mutually exclusive group is defined as a collection of genes in which, for every tumor sample, at most one gene is altered. Conversely, we assume that in a group showing no mutual exclusivity, each gene is altered conditionally independent, given the observation time. In a realistic biological setting however, additional alterations may still provide a small selective advantage to the tumor, rather than none at all, which may lead to the fixation of a genotype with more than one alteration, in a group of genes otherwise perfectly mutually exclusive. Thus, biologically, groups of genes display a continuous range of mutual exclusivity degrees. TiMEx quantifies these degrees exactly, and assesses their significance using a likelihood ratio test. Our procedure for efficient search for mutually exclusive patterns in large datasets consists of three steps (Figure \ref{fig:overview}). We first estimate mutual exclusivity between all possible gene pairs in the dataset. Second, we select as candidates the gene groups in which the significance and degree of mutual exclusivity between each pair of members are high. Third, the candidate groups are statistically tested for mutual exclusivity. 
\indent In simulation studies, we show that TiMEx outperforms the permutation-based method previously introduced by \cite{vandinTest} and the muex model \citep{ewa}. Furthermore, we apply our procedure to four large TCGA studies, two glioblastoma datasets, ovarian \citep{ovarian} and breast (provisional) cancer datasets. On these datasets, we show that TiMEx identifies gene groups with stronger functional biological relevance than the other two methods, while also proposing many new candidates for biological validation. TiMEx doesn't impose any temporal assumptions on the set of biological samples it is applied on. These samples are considered to be independent. Without requiring any previous biological knowledge, our procedure identifies mutually exclusive gene groups of any size, statistically tests and ranks them by their degree of mutual exclusivity. It possesses several advantages over previous methods, including the probabilistic modeling of tumorigenesis as a dynamic process, the novel and intuitive quantification of the degree of mutual exclusivity as a probability, high computational efficiency on large datasets, as well as high sensitivity in detecting low frequently altered genes.

\section*{Methods}
\subsection*{Probabilistic model}
We consider $n$ genes indexed by $N=\{1,2,\ldots,n\}$, whose alteration statuses are represented by the vector of binary random variables $X = \left(X_1,X_2,\ldots,X_n\right)$, recorded at observation time $T_\text{obs} \sim \text{Exp}(\lambda_\text{obs})$. The waiting times to alteration of the $n$ genes are represented by the vector of random variables $T=\left(T_1,T_2,\ldots,T_n\right)$, where $T_i \sim \text{Exp}(\lambda_i), \text{ for all } i \in N$. For a given tumor sample, we refer to an instantiation of $X$, namely $\left(x_1,\ldots, x_n\right)$, as a genotype, where $x_i \in \{0,1\}, \text{ for all } i \in N$. Moreover, for any set of indices $K \subset N$ with cardinality $|K|$, we denote by $g_K$ the genotype for which positions with indices in $K$ are equal to $1$ and positions with indices in $N \setminus K$ are equal to $0$. The presence of an alteration event in a tumor sample signifies both its occurence in one of the tumor cells and its fixation in the measured population, such that the alteration is observed at screening. Let $\mathcal{D}$ denote $M$ independent observations $\left(X^{(1)},\ldots,X^{(M)}\right)$. Each $X^{(j)}=\left(X_1^{(j)}, X_2^{(j)},\ldots,X_n^{(j)}\right)$ denotes the alteration statuses of the $n$ genes in tumor sample $j$, and each $T^{(j)} = \left(T_1^{(j)},T_2^{(j)},\ldots,T_n^{(j)}\right)$  denotes their corresponding waiting times. The binary variables $X_{i}$ are observed, while $T_{i}$ and $T_\text{obs}$ are hidden. We are interested in inferring the degree of mutual exclusivity among the group of $n$ genes. To this end, we compute the likelihood of the data $\mathcal{D}$ under the nested null and mutual exclusivity models introduced below. As the observations are independent, the likelihood of the data under any model $\theta$ is $\displaystyle{ L\left(\theta \mid \mathcal{D}\right) = \prod_{j=1}^M P\left( X^{(j)} \vert \theta \right)}$.
\begin{figure}[h!]
    \centerline {\includegraphics{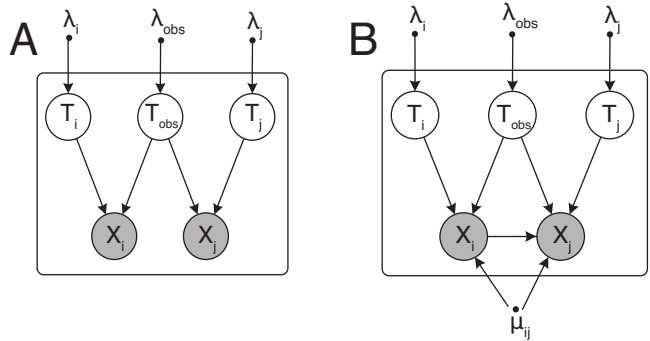}}
    \caption{\footnotesize{Graphical representations of (A) null and (B) mutual exclusivity models for two genes, $i$ and $j$. The observed variables are shaded in gray, while the hidden ones are not. The binary random variables $X_i$ and $X_j$ denote the alteration statuses of the two genes, $T_i$ and $T_j$ are their waiting times, and $T_\text{obs}$ is the observation time, exponentially distributed with corresponding parameters $\lambda$. In the null model, $X_i$ and $X_j$ are conditionally independent given the observation time $T_\text{obs}$, while in the mutual exclusivity model, they also depend on each other via the parameter $\mu_{ij}$, which represents their degree of mutual exclusivity.}}
    \label{fig:models}
\end{figure} 

\subsubsection*{Null Model}
The null model (Figure \ref{fig:models}A), parameterized by $\theta_{\text{Null}} = (\lambda_1,\lambda_2,\ldots, \lambda_n, \lambda_\text{obs})$, assumes that alterations in the $n$ genes are conditionally independent from each other, given the observation time $T_\text{obs}$. The condition for observing an alteration in a gene $i$ is that its corresponding waiting time is shorter than or equal to the observation time: if $T_i \leq T_\text{obs}$, then $X_i=1$, otherwise $X_i=0$. Hence, the dependency between the set of binary variables $X_{i}$ is deterministic and given by the common observation time $T_\text{obs}$.
\\
\indent The genotype $g_{\emptyset}$ is observed if the observation time is shorter than the waiting times of all alterations. Therefore, $T_\text{obs}$ is the minimum of $n+1$ competing exponentials \citep{competingExps}, and
\begin{align}
P\left(g_{\emptyset} \mid \theta_\text{Null}\right) &= P\left(T_\text{obs} < \min_{i \in N} T_i\right) = \frac{\lambda_\text{obs}}{\lambda_\text{obs}+\sum_{i \in N} \lambda_i}
\end{align}
Any genotype $g_K$ with $K \neq \emptyset$ is observed if the waiting times of alterations present in the sample, $T_i, \text{ for all } i \in K$, are shorter than the observation time, and the waiting times of alterations not present in the sample, $T_i, \text{ for all } i \in N \setminus K$, are longer than the the observation time.
\begin{align}
P\left(g_K \mid \theta_\text{Null}\right) &= P\left(\max_{i \in K} T_i \leq T_\text{obs} < \min_{i \in N \setminus K} T_i\right)
\end{align}
The probability that the observation time is shorter than the waiting times of unobserved alterations is not influenced by the specific order between the waiting times of those alterations. Therefore, $P\left(g_K \mid \theta_\text{Null}\right)$ further equals the sum of the probabilities of all possible specific orders of waiting times of observed alterations. Let $S_K = \{(i_{\sigma(1)},i_{\sigma(2)},\ldots,i_{{\sigma(|K|)}}) \mid i_j \in K \text{ for all } j \text{ and } \sigma \in \Sigma_{|K|}\}$ represent the set of all permutations of indices in $K$, where $\Sigma_k$ is the symmetric group of degree $k$. By recursively using the expression of the probability of the minimum of competing exponentials, the probability of observing the genotype $g_K$ is
\begin{align}
P\left(g_K \mid \theta_\text{Null}\right) &= \sum_{\mathclap{\left(i_{1},i_{2},\ldots,i_{|K|}\right) \in S_K}} \quad P\left(T_{i_1} \leq T_{i_2} \ldots \leq T_{i_{|K|}} \leq T_\text{obs} < \min_{i \in N \setminus K} T_i\right) \notag \\
& = \sum_{\substack{ \\ \\ \mathclap{\left(i_{1},i_{2},\ldots,i_{|K|}\right) \in S_K}}} \frac{\lambda_\text{obs}}{\lambda_\text{obs}+\sum_{i \in N \setminus K} \lambda_i} \prod_{j=1}^{|K|} \frac{\lambda_{i_{j}}}{\sum_{l=j}^{|K|} \lambda_{i_{l}} + \lambda_\text{obs} + \sum_{i \in N \setminus K} \lambda_i}
\end{align}
As the observations $\mathcal{D}$ contain no temporal information, the model $\theta_\text{Null}$ is unidentifiable.
\begin{proposition}
The null model $\theta_\text{Null} = \left(\lambda_{1}, \lambda_{2},\ldots, \lambda_{n},\lambda_\text{obs} \right)$ is identifiable only up to $\lamobs$. 
\end{proposition} 
\noindent For the proof, see Supplementary Methods. After setting $\lamobs=1$ (without loss of generality), equivalent to scaling the waiting time rates by $\lambda_\text{obs}$, the reparametrized null model $\theta_\text{Null} = \left(\lambda_{1}, \lambda_{2},\ldots, \lambda_{n} \right)$ becomes identifiable. 

\subsubsection*{Mutual Exclusivity Model}
In the mutual exclusivity model (Figure \ref{fig:models}B), the $n$ genes are assumed to contribute to the same biological function, such that, up to various degrees of mutual exclusivity, only one member is necessary and sufficient to be altered for cancer to progress. An increasing mutual exclusivity interaction in the group directly leads to an increasing fixation probability of a single alteration, corresponding to the gene with the shortest waiting time. The degree of mutual exclusivity of a group of $n$ genes with indices in $N$, denoted by $\mu_{N}$, is the probability that the group is perfectly mutually exclusive. $\mu_{N}$ can also be interpreted as the fractional increase in the fixation probability of the genotypes with a single alteration, when more than one gene in the group were altered before observation time, but, due to the mutual exclusivity interaction between the genes in the group, only the one with the shortest waiting time fixates. The fixation of alterations of further genes is suppressed with probability $\mu_N$. Consequently, $1-\mu_N$ represents the probability of deviating from perfect mutual exclusivity, and for $\mu_{N} \rightarrow 0$, the mutual exclusivity model is reduced to the null model.
\\
\indent The mutual exclusivity model is parametrized by $\theta_{\text{ME}} = (\lambda_1,\lambda_2, \ldots, \lambda_n, \allowbreak \lambda_\text{obs}, \mu_{N})$. The probability of observing the genotype $g_\emptyset$ is the same as in the null model, as the lack of fixated alterations is uninformative for detecting mutual exclusivity,
\begin{align}
P\left(g_{\emptyset} \mid \theta_\text{ME}\right) &= P\left(T_\text{obs} < \min_{i \in N} T_i\right) = \frac{\lambda_\text{obs}}{\lambda_\text{obs}+\sum_{i \in N} \lambda_i}
\end{align}
Any genotype $g_K$ with a single alteration, i.e. with $|K|=1$, can be observed either because $K$ is a mutually exclusive group, or because, by chance, the process of tumorigenessis itself has been observed at the specific point in time when only the alteration with the shortest waiting time in $K$ had fixated. Hence, the probability of observing $g_K$ is the weighted sum of the marginal probability that $T_K$ is simply the shortest waiting time among all waiting times and the probability that the observed alteration pattern happened in the absence of mutual exclusivity interaction between the genes. The first term represents the probability computed under perfect mutual exclusivity and is weighted by $\mu_N$, while the second represents the probability computed under the null model, and is weighted by $1-\mu_N$,
\begin{align}
P\left(g_K \mid \theta_{ME}\right) &= \mu_N P\left(T_{K} < \min_{i \in N \setminus K}\left(T_i,T_\text{obs}\right)\right) + \left(1-\mu_N\right)P\left(T_{K} \leq T_\text{obs} < \min_{i \in N \setminus K} T_i\right) \notag \\
& = \frac{\lambda_{k;k \in K}}{\lambda_\text{obs}+\sum_{i\in N}\lambda_i}  \frac{\lambda_\text{obs}+\mu_N(\sum_{i\in N}\lambda_i-\lambda_{k;k \in K})}{\lambda_\text{obs} + \sum_{i\in N}\lambda_i-\lambda_{k;k \in K}} 
\end{align}
Furthermore, observing any genotype $g_K$ with $|K| >2$, i.e., any genotype with more than one alteration, is considered a deviation from perfect mutual exclusivity. The probability of observing each extra alteration equals the probability that its waiting time is shorter than the observation time, weighted by $1-\mu_N$, the probability of violating perfect mutual exclusivity, 
\begin{align}
P\left(g_K \mid \theta_\text{ME}\right) &= \left(1-\mu_N\right) P\left(\max_{i \in K} T_i \leq T_\text{obs} < \min_{i \in N \setminus K} T_i\right) \notag \\
&= \left(1-\mu_N\right) \sum_{\substack{ \\ \\ \mathclap{\left(i_{1},i_{2},\ldots,i_{|K|}\right) \in S_K}}} \frac{\lambda_\text{obs}}{\lambda_\text{obs}+\sum_{i \in N \setminus K} \lambda_i} \prod_{j=1}^{|K|} \frac{\lambda_{i_{j}}}{\sum_{l=j}^{|K|} \lambda_{i_{l}} + \lambda_\text{obs} + \sum_{i \in N \setminus K} \lambda_i}
\end{align}
\begin{proposition}
\noindent The mutual exclusivity model $\theta_\text{ME} = (\lambda_{1}, \lambda_{2},\ldots, \lambda_{n},$ $\mu_{N}, \lambda_\text{obs})$ is identifiable only up to $\lamobs$.
\end{proposition} 
\noindent For the proof, see Supplementary Methods. Similarly to the null model, after setting $\lamobs=1$ (without loss of generality), the reparametrized mutual exclusivity model $\theta_\text{ME} = \left(\lambda_{1}, \lambda_{2},\ldots, \lambda_{n}, \mu_{N} \right)$ becomes identifiable.

\subsubsection*{Parameter estimation and testing}
The maximum likelihood estimates of all parameters are obtained by setting to zero the corresponding first derivative of the observed log likelihood, numerically approximated using the gradient projection method \citep{optimization} (Figures S1 and S2). An exception is the case $n=2$, which allows for an analytical solution for $\theta_\text{Null}$ (the estimates are given in Supplementary Methods). 

\begin{proposition}
If $n=2$, then there exists a closed-form solution for the maximum likelihood estimates of $\lambda_1$ and $\lambda_2$ under the null model $\theta_\text{Null}$.
\end{proposition}

\noindent To test for any degree of mutual exclusivity interaction among the $n$ genes, we are testing the alternative hypothesis $\mu_{N} \neq 0$ versus the null hypothesis $\mu_{N}=0$. The logarithm of the ratio of the two likelihoods computed for the maximum likelihood estimates is $\chi^2$ distributed with one degree of freedom \citep{neymanLRT}. The likelihood ratio test statistic is well behaved, as under the null hypothesis, the p-values are uniformly distributed (Figure S3). 

\subsection*{Overall procedure and computational complexity}
Our procedure consists of three steps. Given a large dataset of $k$ genes, we first test all $\binom{k}{2}$ pairs for mutual exclusivity, estimating $\theta_\text{Null}$ and $\theta_\text{ME}$ for each pair. The computational complexity of this step is $\mathcal{O}(k^2)$. Second, we construct an undirected graph in which genes are vertices and an edge is drawn between any pair $(i,j)$ if, for chosen thresholds $p_\text{pair}$ and $\mu_\text{pair}$, the p-value $p_{ij} \leq p_{\text{pair}}$, and the degree of mutual exclusivity $\mu_{ij} \geq \mu_{\text{pair}}$. The thresholds are chosen based on the sensitivity and specificity levels to which they correspond, as assessed in simulated data. Further, we produce group candidates by listing all maximal cliques in the constructed graph. To this end, we use the Bon-Kerbosch recursive backtracking algorithm \citep{Bron-Kerbosch}. The upper bound on the running time of the Bon-Kerbosch algorithm is $\mathcal{O}(3^{k/3})$. However, in practice, it is highly efficient \citep{maxCliques}. Finally, we test the candidate groups for mutual exclusivity, and select the ones for which the Bonferroni corrected p-value is lower than a chosen cutoff. Due to the cubic complexity of matrix inversion (in standard Gauss-Jordan elimination) employed by the numerical optimization routine \citep{optimization}, the complexity of the last step has an upper bound of $\mathcal{O}(s_q q^3)$, where $q$ is the maximal identified clique size, and $s_q$ the number of such cliques of this size.

\begin{figure}
	\centerline  {\includegraphics{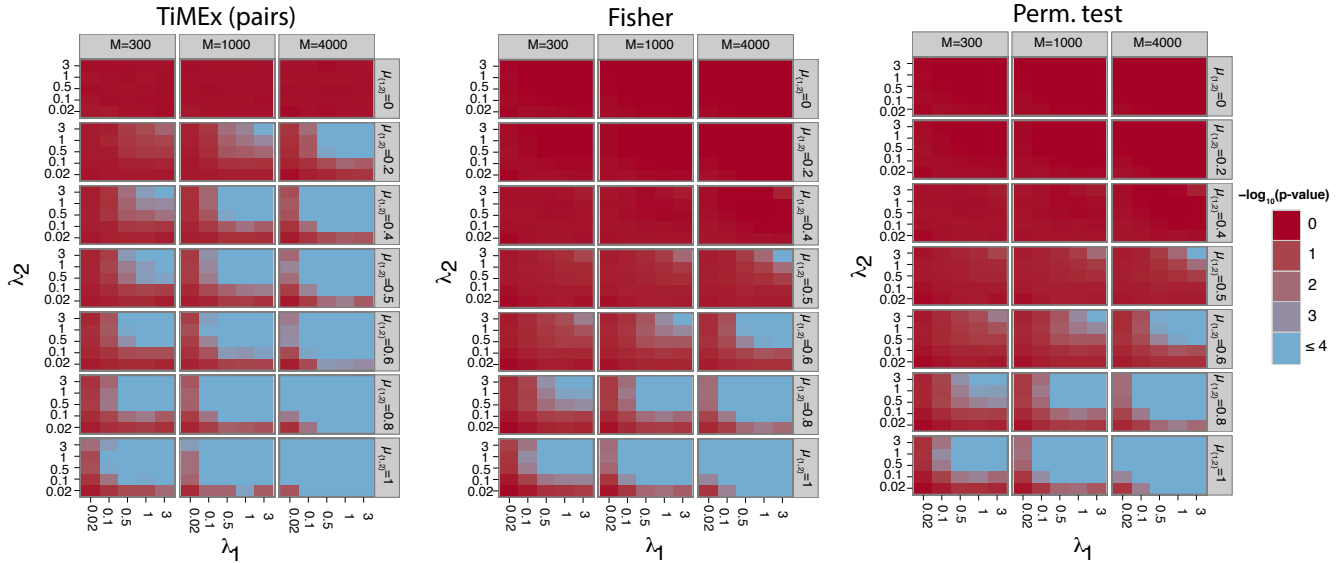}}
	\caption{\footnotesize{Mean p-value (over 100 simulation runs) for TiMEx, Fisher's exact test, and the permutation test in \cite{vandinTest}, for different sample sizes $M$, pairwise degrees of mutual exclusivity $\mu_{\{1,2\}}$, and waiting time rates determining the marginal frequencies of the two genes, $\lambda_1$ and $\lambda_2$. TiMEx is highly sensitive in detecting mutual exclusivity, and outperforms the other two tests, which only start detecting mutual exclusivity for $\mu_{\{1,2\}} \geq 0.5$. The detection capacity increases with increasing values of $\mu_{\{1,2\}}$, $M$, $\lambda_1$, and $\lambda_2$.}}
     \label{fig:simPvalsPairs}
\end{figure}

\section*{Results}
\subsection*{Simulations}
We assessed the behavior and performance of TiMEx on simulated data, by varying the waiting time rates of the genes, the degrees of mutual exclusivity of the group, and the sample sizes. Specifically, the values of the sample size $M$ were $300$, which is similar to the size of the ovarian cancer dataset, $1000$, similar to the size of the breast cancer dataset, and $4000$, which is a realistic estimate for the size of genomic datasets in the near future. The degrees of mutual exclusivity used for simulations were  $\mu \in \{0,0.2,0.4,0.5,0.6,0.8,1\}$. We compared the ability of mutual exclusivity detection of TiMEx, for both pairs and groups, with a previously introduced permutation-based method in \cite{vandinTest} (ran with 1000 permutations). For the tests on pairs, TiMEx was further compared with one-sided Fisher's exact test for contingency tables, testing whether the number of double mutants is significantly lower than expected under independence. For the test on larger groups, TiMEx was additionally compared with muex \citep{ewa}, a previously introduced statistical model for detecting mtuually exclusive groups. In a power analysis, we investigated how the sensitivity and specificity of our procedure are influenced by the thresholds on significance and mutual exclusivity degree, $p_\text{pair}$ and $\mu_\text{pair}$.

\subsubsection*{Test performance for pairs and groups}
For simulating mutually exclusive gene pairs, we used $\lambda_1$, $\lambda_2$ $ \in \{0.02,0.1,0.5,1,3\}$, corresponding to marginal frequencies of the two genes ranging from $2\%$ to $75\%$ in the null model and $0.5\%$ to $75\%$ in the mutual exclusivity model with $\mu_{\{1,2\}}=1$ (Tables S1 and S2). We performed 100 simulation runs, detected pairwise mutual exclusivity with the three tests, and recorded the mean p-value (Figure \ref{fig:simPvalsPairs}). In the case where $\mu_{\{1,2\}}=0$, corresponding to lack of mutual exclusivity, all three tests do not reject the null hypothesis, with a p-value close to 1, for all tested combinations of frequencies and sample sizes. TiMEx is the only test that starts detecting mutual exclusivity from the first non-zero value of $\mu_{\{1,2\}}$ in the chosen simulation set, however with reduced performance for small sample size and small frequencies of both genes. The detection capacity increases with increasing values of $\mu_{\{1,2\}}$, $M$, $\lambda_1$, and $\lambda_2$. For example, for a chosen significance level of $0.05$ and a sample size of $M=4000$, TiMEx detects the gene pairs as being mutually exclusive for any value of $\mu_{\{1,2\}} \geq 0.4$ and for any $\lambda_1$, $\lambda_2 \geq 0.1$. For higher marginal frequencies such as, for example, corresponding to $\lambda_1, \lambda_2 \geq 0.5$, we can detect mutual exclusivity of degree $\mu_{\{1,2\}} \geq 0.4$ for sample sizes as low as $M=300$. By contrast, Fisher's exact test and the permutation test in \cite{vandinTest}, while performing highly similarly to each other, detect no mutual exclusivity for $\mu_{\{1,2\}} <0.5$. Moreover, for $\mu_{\{1,2\}} \geq 0.5$, their detection ability is much reduced compared to TiMEx. The null model used in Fisher's exact test is a classical independence model, while the waiting times in our null model are not statistically independent, even with fixed rate of the observation time $\lambda_\text{obs}$.
\begin{figure}[t!]
	\centerline  {\includegraphics{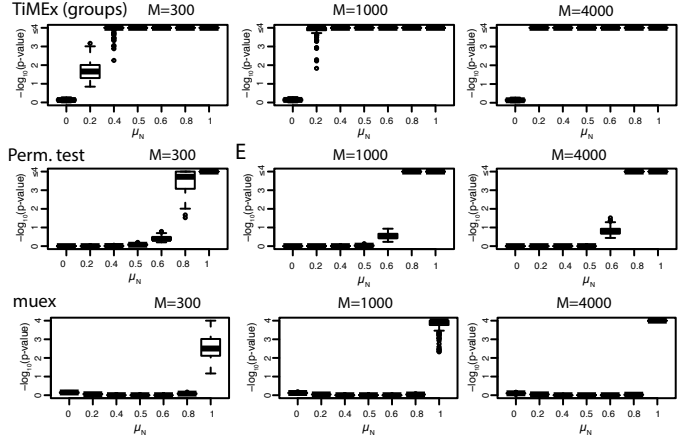}}
	\caption{\footnotesize{Summary pvalue (over 100 simulation runs and 10 simulated groups of size 5) of TiMEx, the permutation test in \cite{vandinTest} and muex \citep{ewa}, for different sample sizes $M$ and degrees of mutual exclusivity $\mu_N$. TiMEx is highly sensitive in detecting mutual exclusivity, and outperforms both the other two methods. The permutation test only detects mutual exclusivity for $\mu_N \geq 0.8$ and outperforms muex, which only detects pure mutualy exlusivity ($\mu_N=1$). The detection ability of TiMEx improves with increasing values of $M$ and $\mu_N$.}}
     \label{fig:simPvalsGroups}
\end{figure} 
\\
\indent For simulating mutually exclusive groups, we fixed the group size to $5$ and produced $10$ different groups by uniformly sampling waiting time rates with values between $0.01$ and $1$, which corresponds to an expected alteration frequency of $14\%$ (Table S3). We performed 100 simulation runs, detected mutual exclusivity with TiMEx, the permutation test, and the muex model, and summarized the p-value over both different simulated groups and simulation runs (Figure \ref{fig:simPvalsGroups}). Similarly to the case of pairs, the detection ability of TiMEx increases with increasing sample size and degree of mutual exclusivity. For a significance level of $0.05$, we detect mutual exclusivity for almost all tested sample sizes and mutual exclusivity degrees, with the exception of small sample size $M=300$ and low degree of mutual exclusivity $\mu_N=0.2$. For the highest sample size, $M=4000$, TiMEx is very sensitive in detecting mutual exclusivity for any tested positive degree, with a mean p-value $\leq 10^{-4}$. By contrast, the permutation test only starts detecting mutual exclusivity for $\mu_N \geq 0.8$, however outperforming muex, which only detects pure mutual exclusivity ($\mu_N=1$). On data simulated using $\mu_N=0$, all three tests do not reject the null hypothesis with mean p-value $> 0.6$.
\\
\indent In addition to assessing the detection ability of TiMEx and other methods on data simulated from TiMEx, we conducted simulations on datasets generated more generally. We generated groups of $n=5$ mutually exclusive genes by varying the sample size as before, the coverage, i.e. the percentage of patients which have at least one gene altered, among $\{0.2,0.4,0.6,0.8\}$, and the probability of passenger alterations among $\{0.001,0.01,0.1\}$. Depending on the coverage, before adding noise, at most one gene was altered in each patient, which rendered the group perfectly mutually exclusive. Passenger mutations were further added to each patient with the chosen probability. On all datasets, TiMEx outperforms the permutation test and muex, and always records lower p-values (ranking not shown for p-values lower than $10^{-10}$) (Figure S4). All three methods perform better with increasing sample size, increasing coverage, and decreasing passenger probability, and, for most of the tested values, they significantly detect the group as mutually exclusive. For TiMEx, we also estimated the degree of mutual exclusivity $\mu_N$ corresponding to the generated groups (Figure S5). For the very low passenger probability of $0.001$, the inferred degree of mutual exclusivity is $1$, as the expected number of passenger mutations per dataset is very low, especially for small sample sizes. The lowest inferred degree of mutual exclusivity is $0.7$, corresponding to small coverage and small sample size. The estimated $\mu_N$ increases with increasing coverage and decreasing passenger probability, and the estimation improves with increasing sample size.

\subsubsection*{Power analysis}
For assessing the true and false positive rates of our procedure, we constructed 100 datasets consisting of two groups: a group of size 3 simulated from the mutually exclusive model, and a group of size 9 simulated from the conditionally independent model, with $\lambda$ values sampled uniformly between $0.01$ and $1$ (Table S3). We tested all pairs with TiMEx, detected maximal cliques as candidates, and evaluated them with TiMEx. We considered a detected group to be mutually exclusive if its Bonferroni corrected p-value was lower than 0.1. We computed the true positive rate by counting a single time, among the detected mutually exclusive groups of size at least 3, all edges only connecting two genes part of the true mutually exclusive group, and normalizing by the number of all possible such edges. Similarly, we computed the false positive rate by counting a single time, among the detected mutually exclusive groups of size at least 3, all edges not connecting two genes part of the true mutually exclusive group, and normalizing accordingly. TiMEx performs generally very well in reconstructing the implanted mutually exclusive group (Figure S7). The highest effect in increasing the true positive rate and decreasing the false negative rate was given by increasing the mutual exclusivity degree of the simulated group. For values of the threshold $\mu_\text{pair} \geq 0.5$, the false positive rate was often set to $0$, and most of the times reduced by at least 75$\%$ as compared to the case when $\mu_\text{pair}=0.2$. However, for small degrees of mutual exclusivity of the simulated group, the true positive rate was often reduced simultaneously with the false positive rate.
\\
\indent Additionally, on the same simulated datasets, we analyzed how often the true mutually exclusive group is also the top ranked group by corrected p-value (Figure S8). The threshold $\mu_\text{pair}$ largely impacts the detection performance, while the impact of $p_\text{pair}$ is neligible. For high degrees of mutual exclusivity of the true group, $\mu_N \geq 0.8$, the real group is either top ranked, or a strict subset of the top ranked one, depending on the value chosen for $\mu_\text{pair}$. Optimal performance is achieved for $\mu_\text{pair} = 0.5$, corresponding to a percentage of between $50\%$ and $100\%$ of datasets for which the true group is top ranked. Moreover, if the true group is perfectly mutually exclusive ($\mu_N=1$), it is top ranked in more than $90\%$ of the datasets for medium values of $\mu_\text{pair}$, for any value of $p_\text{pair}$, and for medium sample sizes. For lower values of $\mu_{pair}$, the true group is a strict subset of the top ranked group. For low degrees of mutual exclusivity, either no significant groups are identified, or the top ranked group is not the real group. The detection power improves with increasing sample size and increasing degree of mutual exclusivity $\mu_N$. 

\subsection*{Biological datasets} 
We ran our procedure on four biological datasets: the two glioblastoma datasets preprocessed by muex \citep{ewa} and Multidendrix \citep{multidendrix}, and two datasets downloaded from TCGA and preprocessed as explained in Section S2.3: breast cancer and ovarian cancer. Our main interest was detecting gene groups with average or high degree of mutual exclusivity and minimizing the false positive rate, while maintaining the true positive rate at a high level. Therefore, based on the sensitivity and specificity estimates in simulated data (Figures S7 and S8), we set $\mu_\text{pair} = 0.5$ and $p_\text{pair} = 0.01$ for the four datasets.  A detected group was considered significantly mutually exclusive if its Bonferroni-corrected p-value (q-value) was less than 0.1 (Figure S9). In order to test the stability of the identified groups, we subsampled the set of patients at different frequencies: $30\%$, $50\%$, and $80\%$, and repeated the procedure 100 times, reporting how often each group is still identified as mutually exclusive (Tables S4-S15). Among the identified groups of any size, we further computed the most stable subgroups. For mutually exclusive groups with high enough alteration frequencies, higher stability indicates stronger mutual exclusivity support in the data. For each group size, we tested the first $10$ groups ranked by q-value for pathway enrichment with WebGesalt \citep{webgestalt} on the Pathways Commons dataset \citep{pathwaycommons}, and reported all significantly enriched pahtways for a BH-corrected p-value threshold of 0.01. On all four datasets, we compared our results with two other methods: Multidendix, an algorithm based on the permutation test we used for comparison on simulated data, and muex (Tables S31-S36). Section S1 discusses the mutually exlcusive groups identified in the two glioblastoma datasets. 

\subsubsection*{Mutual Exclusivity in Breast Cancer}
\begin{figure}
	\centerline  {\includegraphics{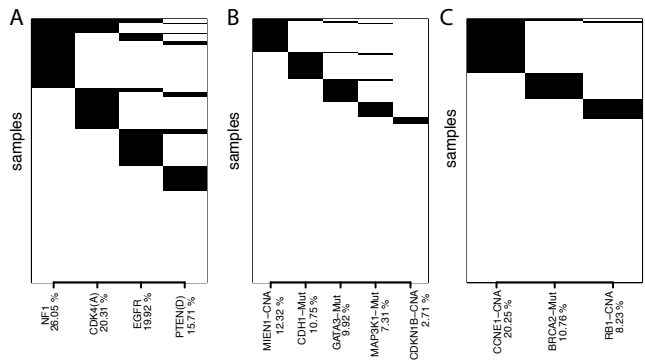}}
	\caption{\footnotesize{The alteration frequencies of selected mutually exclusive groups, as identified by our procedure. The horizontal axis displays the members of each group, together with their relative frequency in the dataset, as well as their alteration type (\textit{Mut} for mutation, and \textit{CNA} for copy number aberration). A black line is drawn whenever an alteration is present in a sample. (A): the group consisting of the deletion of \textit{PTEN}, the amplification of \textit{CDK4}, and the point mutations EGFR and NF1 (q-value 3e-10) was the most stable among the groups of largest size ($89\%$ recovery at subsampling $80\%$ of the patients) identified by TiMEx in the glioblastoma dataset used by Multidendrix in \cite{multidendrix}. (B): the group consisting of the point mutations of \textit{CDH1}, \textit{MAP3K1}, \textit{GATA3}, and the copy number aberrations of \textit{CDKN1B}, \textit{MIEN1} (q-value 1e-23) was the most significant group of largest size identified by TiMEx in the breast cancer dataset. (C): the group consisting of the point mutation of \textit{BRCA2} and the copy number aberrations of \textit{RB1} and \textit{CCNE1} (q-value 5e-09) was the most significant group of largest size identified by TiMEx in the ovarian cancer dataset.}}
     \label{fig:bioGroups}
\end{figure}

In the breast cancer dataset (Figure \ref{fig:bioGroups}), we found $63$ groups of size two, $416$ groups of size three, $96$ groups of size four, and $10$ groups of size five. Since all the $10$ largest groups contained one gene with frequency less than $10\%$, these groups were highly unstable to subsampling, even if they corresponded to functionally related collections of genes (Table S13). The first three groups with the lowest q-value consisted of the point mutations of the tumor suppressors \textit{CDH1}, \textit{GATA3}, \textit{MAP3K1}, the copy number aberration of \textit{CDKN1B}, which belong to pathways including PI(3)K, mTOR, PDGF receptor signaling network, or EGF receptor (ErbB1) signaling, and the copy number aberration of one of the following genes: \textit{MIEN1}, \textit{PPP1R1B}, or \textit{ERBB2}. \textit{MIEN1} is an oncogenic protein, whose overexpression functionally enhances migration and invasion of tumor cells via modulating the activity of the PI(3)K pathway \citep{mien1}, providing evidence for the functional relation between these genes. Moreover, the \textit{PPP1R1B-STARD3} chimeric fusion transcript was shown to activate the PI(3)K/AKT signaling pathway and promote tumorigenesis \citep{ppp1r1b}, while \textit{ERBB2} is an oncogene that also belongs to the PI(3)K and mTOR pathways. The next two mutually exclusive groups of size five included the same three point mutations, the copy number aberration of \textit{MIEN1}, and the copy number aberrations of either \textit{B4GALNT3}, which has no known functional role in breast cancer, or \textit{GRB7}, which is part of the \textit{Common group of pathways}. The first groups of size four with lowest q-value consisted of the point mutations of \textit{CDH1}, \textit{MAP3K1}, \textit{TP53}, and \textit{GATA3}, and was entirely mapped to the \textit{Common group of pathways}, as well as to the CDC42 signaling events pathway (Table S12). The second and the third group included, instead of the \textit{GATA3} point mutation, the copy number aberration of either \textit{TUBD1} or \textit{INTS4}. Even though strong evidence of association for these two genes and the group of three point mutations exists in the data, \textit{TUBD1} and \textit{INTS4} have no known functional role in cancer. The subgroups with highest subsampling stability (Tables S22-S24) consisted of genes with known functional involvement in cancer, such as \textit{GATA3}, \textit{PIK3CA}, or \textit{PTEN}.
\\
\indent We separately ran our procedure on the subset consisting of $507$ samples annotated as distinct breast cancer subtypes (Tables S27-S30). Some of the top ranked mutually exclusive relations identified based on the entire dataset were also identified based on the subsets of data belonging to Her2, LuminalA, and LuminalB subtypes. None of the alterations identified in the top ranking groups were specitic to the Basal subtype (Table S27). For example, the connections between one of the point mutations of \textit{PIK3CA} or \textit{CDH1}, or the copy number aberration of \textit{PTEN}, and the copy number aberrations of one of \textit{ERBB2}, \textit{GBR7}, \textit{MIEN1}, \textit{PNMT}, or \textit{PPP1R1B} were also mutually exclusive in the Her2 subtype (Table S28). Similarly, the mutually exclusive group consisting of the point mutations \textit{MAP3K1}, \textit{GATA3}, and \textit{TP53} was identified in the LuminalA subtype (Table S29), while the group including the point mutations \textit{PIK3CA}, \textit{TP53}, and \textit{GATA3} was LuminalB subtype-specific (Table S30). Also, \textit{TUBD1}, a gene part of mutliple groups, was mutually exclusive with the point mutation of \textit{MAP3K1} in LuminalA, and with the point mutations of \textit{PIK3CA} and \textit{TP53} in LuminalB.
\\
\indent We ran Multidendrix on the breast cancer dataset, using $\alpha=2.5$ (as suggested in \cite{multidendrix}), $t=4$, and a range of $k_\text{max}$ values (Table S34). Multidendrix identified with highest weight the core group including the point mutations of \textit{TP53}, \textit{GATA3}, and \textit{MAP3K1}, however in the same group as the point mutations of \textit{CTCF} and \textit{PLXNB2}, which are not part of any of the known functional pathways. On the contrary, TiMEx identified these three point mutations in a common module with the point mutation of \textit{CDH1}. Similarly, the next two modules ordered by weight only contained three genes in known functional pathways, as assessed by WebGestalt, on the Pathway Commons database (data not shown). The fourth module identified by Multidendrix contained no signifincat pathways. muex did not scale to the size of dataset, and  none of the top $30$ groups of any size identified by TiMEx were found significantly mutually exclusive by muex's statistical test.

\subsubsection*{Mutual Exclusivity in Ovarian Cancer}
In the ovarian cancer dataset (Figure \ref{fig:bioGroups}), we identified $24$ mutually exclusive groups of size two and $24$ groups of size three. The top ranked group of size three (Table S15) included three genes part of the FOXM1 transcription factor network, and involved in cell cycle regulation, recently shown to play a major role in the progression of ovarian cancer \citep{ovarian}: the copy number aberrations of the tumor suppressor gene \textit{RB1} and the oncogene \textit{CCNE1}, and the point mutation of the tumor suppressor gene \textit{BRCA2}. The subgroup consisting of \textit{RB1} and \textit{CCNE1} was also the most stable to subsampling (Tables S26). Among the top five groups of size $3$, the one which was most stable to subsampling included core members of the ATM pathway: the point mutations of \textit{BRCA1} and \textit{BRCA2}, and the copy number aberration of \textit{CCNE1}. These two modules have also been previously identified as mutually exclusive by MEMo, an algorithm for detecting mutually exclusive groups \citep{memo}. The following top scoring groups of size three included the copy number aberrations of \textit{MYC} and \textit{CCNE1}, two members of cell cycle regulation pathways involved in the G1/S phase transition, also identified by MEMo, together with the copy number aberration of one gene with yet unknown functional role in ovarian cancer: \textit{WNK1}, \textit{NINJ2}, or \textit{B4GALNT3} (also identified in breast cancer). The top ranked mutually exclusive pair, which was also the most stable (identified $54\%$ of the times when subsampling $80\%$ of the patients) included \textit{KRAS} and \textit{TP53} point mutations, which are part of the p75 NTR receptor-mediated signalling pathway (Table S14). The second mutually exclusive pair included the point mutations of \textit{TP53} and \textit{RB1}, both part of the TGFBR and p53 pathways.
\\
\indent We ran Multidendrix on the ovarian cancer dataset, using $t=4$ and a range of $k_\text{max}$ values (Table S35). The groups identified by TiMEx and Multidendrix showed a high overlap. For example, the top ranking groups identified by TiMEx, i.e. the pair including the point mutations of \textit{TP53} and \textit{KRAS}, and the group including the copy number aberrations of \textit{RB1} and \textit{CCNE1}, together with the point mutation of \textit{BRCA2}, were also identified by Multidendrix. Moreover, subsets of most of the group members that Multidendrix identified for e.g. $k_\text{max}=5$ were identified by TiMEx as groups of size three, such as the point mutations of \textit{BRCA1} and \textit{BRCA2} and the copy number aberration of \textit{EPHX3}. Even though muex did not scale to exhaustively analyze the dataset for groups, 14 of the pairs identified by TiMEx and 4 of the groups of size three were found to be significant by muex (Table S36). Almost all the pairs included either the point mutation of \textit{BRCA2} or the copy number aberration of \textit{NF1}, while the larger groups included genes mapping to relevant pathways, among which many had also been identified by Multidendrix. The reason why these groups are also found to be mutually exclusive by muex is the fact that the alteration frequencies of their members are balanced.

\section*{Discussion}
We have introduced TiMEx, a probabilistic generative model for detecting mutual exclusive patterns of various degrees across carcinogenic alterations, and an efficient multistep procedure for identifying all mutually exclusive groups in large datasets. TiMEx is the first method that describes the mutual exclusivity property as a consequence of a dynamic process in time. Unlike previous \textit{de novo} approaches, TiMEx infers functional relations between genes based on an underlying temporal representation of the process of gene alteration in tumorigenesis. Moreover, TiMEx is a probabilistic generative model, providing a natural way of rigurously quantifying the degree and significance of mutual exclusivity of a group of genes. Furthermore, to the best of our knowledge, TiMEx is the first method inferring a continuous range of mutual exclusivity degrees. Biologically, the small, but observable, increase in tumor fitness due to multiple alterations in a group of functionally related genes supports the hypothesis that mutual exclusivity occurs at various degrees, as opposed to a binary classification \citep{memo}. Unlike most other approaches, TiMEx does not explicitly impose constraints on frequencies of alterations, in order to identify them as mutually exclusive. Our procedure detects both high frequent and very low frequent alterations, only based on the temporal relation between them. Finally, it identifies all mutually exclusive gene groups of various, not pre-defined sizes, and performs highly efficiently on large datasets.
\\
\indent  TiMEx is however still a simplified representation of carcinogenesis. Given a particular order between the waiting times of the genes and the observation time, the probability of violating mutual exclusivity, $1-\mu_N$, is independent of how many, or which alterations are in a group. One natural extension of TiMEx would be to consider an incremental penalty for additional point alterations violating perfect mutual exclusivity, hence increasing the probability of being in a non mutually exclusive state with increasing number of violating alterations. Additionally, even if highly efficient, the search for mutually exclusive gene groups is heuristic, and depends on the thresholds $p_\text{pair}$ and $\mu_\text{pair}$. With overly stringent thresholds, too few candidates would be proposed, while using overly permisive thresholds would lead to selecting as candidates a vast number of subsets, making the procedure intractable. To address this, we propose setting the thresholds following the desired sensitivity-specificity tradeoff as assessed in simulations. Moreover, the functional role in tumorigenesis that specific genes might have can be analyzed in higher detail by simply including different point mutations of the same gene as separate alterations. 
\\
\indent The exponential distribution, used for modeling the waiting times to alterations and the observation time, is a typical choice to describe waiting times \citep{moritzniko}, both due to its generality and to its mathematical convenience. While the exponential distribution is the simplest model for system failure time, other families of distributions for modeling the observation time can be readily integrated into our mathematical framework, with nevertheless the cost of more involved mathematical formulas. For example, using the Weilbull distribution provides a supporting assumption in modeling cancer progression due to the fact that the instantaneous probability of occurrence of an event changes with time. However, the superiority of such choices would need to be evaluated in future applications. Another extension of TiMEx is renouncing to the independence assumption at the level of observations, and applying our procedure to large-scale time series data of tumor progression. Once this type of data becomes available, TiMEx will facilitate a more detailed understanding of pathways involved in tumor progression. 
\\
\indent In simulation studies, TiMEx outperforms previous methods for detecting mutual exclusive groups, showing high sensitivity even at low degrees of mutual exclusivity and scaling very well to sample sizes of several thousands tumors, which is expected to be soon reached by cancer genome sequencing studies. On biological datasets, most of the top ranked mutually exclusive groups identified by TiMEx have stronger functional biological relevance than the groups identified by previous methods. In conclusion, results on both simulated and biological data clearly indicate that TiMEx is not only theoretically justified by its biological and probabilistic foundation in describing tumorigenesis as a generative process of mutually exclusive alteration patterns, but is also efficiently and fruitfully applicable in practice.
\vspace{-1.5mm}
\section*{Acknowledgement}
The authors would like to thank Hesam Montazeri, David Seifert, and Jack Kuipers for useful discussions and suggestions.

\bibliographystyle{plainnat}
\bibliography{bibME}

\clearpage
\includepdf[pages={-}]{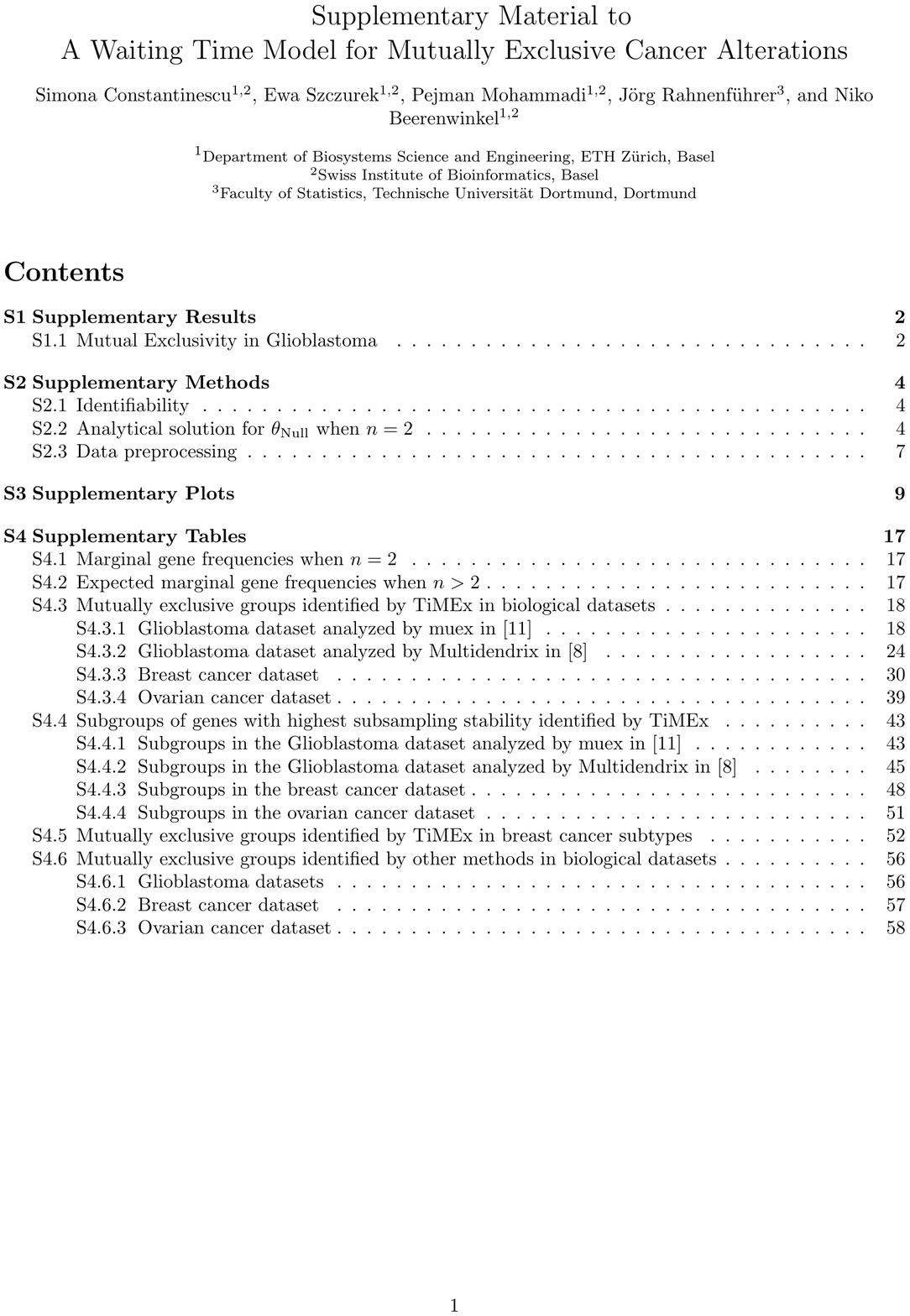}

\end{document}